\documentstyle[psfig]{l-aa} 
\def\pBoxtimes{{\ooalign{\hfil\raise.2ex\relax
\hbox{$\times$}\hfil\crcr\hbox{$\Box$}}}}

\def\pBoxcirc{{\ooalign{\hfil\raise.2ex\relax
\hbox{$\circ$}\hfil\crcr\hbox{$\Box$}}}}

\def\pDiatimes{{\ooalign{\hfil\raise.2ex\relax
\hbox{$\times$}\hfil\crcr\hbox{$\Diamond$}}}}

\def\pBoxplus{{\ooalign{\hfil\raise.2ex\relax
\hbox{$+$}\hfil\crcr\hbox{$\Box$}}}}
\newcommand{\araa}{ARA\&A}   
\newcommand{\aj}{AJ}         
\newcommand{\aaa}{A\&A}      
\newcommand{\aas}{A\&AS}     
\newcommand{\apj}{ApJ}       
\newcommand{\apjs}{ApJS}     
\newcommand{\apss}{Ap\&SS}   
\newcommand{\mnras}{MNRAS}   
\newcommand{\pasp}{PASP}     
\newcommand{\qjras}{QJRAS}   
 
\newcommand{\Msun}{\,\mbox{M}_{\odot}}

\newcommand{\scrm}[1]{\mbox{\scriptsize\rm #1}}
\newcommand{\tinm}[1]{\mbox{\tiny\rm #1}}

\def\lmean{\mathopen{<}}
\def\rmean{\mathclose{>}}
\newcommand{\etal}{et al.\ }

\newcommand{\eg}{e.g.,\ }

\newcommand{\Mzon}{M$_{\odot}$}
\newcommand{\Lzon}{L$_{\odot}$}

\newcommand{\Av}{$A_{\scrm{V}}$}

\newcommand{\kms}{\mbox{km s$^{-1}$}}
\newcommand{\s}{\mbox{s$^{-1}$}}
\newcommand{\ergs}{\mbox{erg s$^{-1}$}}
\newcommand{\ergcms}{erg s$^{-1}$ cm$^{-2}$} 
\newcommand{\Mpc}{Mpc$^{-1}$} 
\newcommand{\Ha}{H$\alpha$}

\newcommand{\SII}{[{\sc S$\,$ii}]}
\newcommand{\NII}{[{\sc N$\,$ii}]}

\newcommand{\multi}{\multicolumn}

\begin{document}

\title{X-ray emission, optical nebulosity and dust in 
early-type galaxies.} 
\subtitle{I.\ The dusty nebular filaments in NGC 5846\thanks{Based on
 observations collected at the European Southern Observatory, La Silla,
 Chile}} 

\author{P.\ Goudfrooij\inst{1,2}  \and G.\ Trinchieri\inst{3,4}}

\offprints{P.\ Goudfrooij {\sf (goudfroo@stsci.edu)}}

\institute{
 Space Telescope Science Institute, 3700 San Martin Drive, Baltimore,
 MD 21218, U.S.A.
\and 
 Affiliated to the Astrophysics Division, Space Science Department,
  European Space Agency
\and 
 Max-Planck-Institut f\"ur extraterrestrische Physik,
 Giessenbachstrasse 1, D-85740 Garching bei M\"unchen, Germany
\and 
 Osservatorio Astronomico di Brera, Via Brera 28, I-20121 Milano, Italy
}

\thesaurus{03(11.05.1; 11.06.2; 11.09.1 NGC 5846; 11.09.4; 11.19.6)} 

\date{Received September 1 / Accepted ..., 1997}

\maketitle

\markboth{P.\ Goudfrooij and G.\ Trinchieri:\ The dusty nebular
filaments in NGC 5846}{P.\ Goudfrooij and G.\ Trinchieri:\ The dusty
nebular filaments in NGC 5846} 

\begin{abstract}
We present new optical imagery of the
X-ray bright elliptical galaxy NGC 5846, the dominant galaxy of a
small group of galaxies. A filamentary dust lane with
a dust mass of $\sim$\,7  $10^3$ \Mzon\ is detected in the central few
kpc of NGC 5846. The optical extinction properties of the dust
features are consistent with those of dust in our Galaxy. The
morphology of the dust  features are strikingly similar to that
observed for the optical nebulosity {\it and\/} the X-ray emission. A
physical connection between the different phases of the interstellar
medium therefore seems likely. 

We discuss three different options for the origin of the dusty nebular
filaments: Condensation out of a cooling flow, mass loss of late-type
stars within NGC~5846, and material donated by a small neighbouring
galaxy. We conclude that the dust as well as the optical nebulosity
are most likely products of a recent interaction with a small,
relatively gas-rich galaxy, probably of Magellanic type. 

Dust grains in the dust lane are destroyed by sputtering in the hot,
X-ray-emitting gas in $\la 10^7$ yr, which is 
shorter than the crossing time of a (small) galaxy through the central
5 kpc of NGC~5846. This
indicates that the dust must be replenished to be consistent with the
observed dust mass, at a rate of $\sim$\,10$^{-3}$ \Mzon\ yr$^{-1}$. 
We argue that this replenishment can be achieved by evaporation of
cool, dense gas cloudlets that were brought in during the
interaction. The evaporation rate of cool gas in NGC~5846 is 
consistent with the ``mass deposition rate'' derived from X-ray
measurements.  

The energy lost by the hot gas through heating of dust grains and
evaporation of cool gas clouds in the central few kpc of NGC~5846 is
adequately balanced by heat 
sources: transport of heat by electron conduction into the core of the 
X-ray-emitting gas and loss of kinetic energy of the in-falling
galaxy. There does not seem to be a need to invoke the presence
of a ``cooling flow'' to explain the X-ray and optical data. 




\end{abstract}

\section{Introduction}
\label{intro}
Elliptical galaxies have long been considered to be inert stellar
systems, essentially devoid of interstellar matter. However, our
understanding of the nature of the interstellar medium (ISM) in 
elliptical galaxies has undergone a radical change from this consensus that
prevailed only a dozen of years ago. Recent advances in instrumental
sensitivity across the electro-magnetic spectrum have revealed the
presence of a complex, diverse ISM in elliptical galaxies.

The ISM of luminous ($M_B \la -21.0$) elliptical galaxies is generally
dominated (in mass) by hot ($T \sim 10^7$ K), X-ray-emitting gas
($10^9 - 10^{11} \Msun$\footnote{H$_0$ = 50 \kms\ \Mpc\ is assumed
throughout this paper}; e.g., Canizares, Fabbiano \& Trinchieri
\cite{cft87}).  Relatively small, varying quantities of cool neutral
gas ($\la 10^7 \Msun$; cf.\ Lees \etal \cite{lees+91}; Wiklind, Combes
\& Henkel \cite{wikl+95}), dust ($10^4 - 10^6 \Msun$; e.g., Knapp
\etal \cite{knapp+89}; Goudfrooij \& de Jong \cite{gdj95}), and
optical nebulosity ($10^3 - 10^5 \Msun$; e.g., Buson \etal
\cite{buson+93}; Goudfrooij \etal \cite{paul+94b}; Macchetto \etal
\cite{duccio+96}) have also been detected. Hence, all ISM components
known to exist in spiral galaxies are now accessible in elliptical
galaxies as well (although in rather different proportions).  Unlike
the situation in spiral galaxies however, it 
still remains  unclear ---and controversial--- what the correct
description of that ISM is, i.e., what the nature, origin and fate of
the different components of the ISM are, and what the dominant
process is that dictates the interplay between them. 

A number of theoretical concepts have been developed for the secular
evolution of the ISM of luminous elliptical galaxies. One of the 
currently most popular concepts is the ``cooling flow'' scenario
(recently reviewed by Fabian, \cite{fabi94}):\ The global dynamics of 
the hot gas present in X-ray bright elliptical galaxies is controlled by
the relative efficiencies of heating and cooling processes. The
cooling time of the gas decreases strongly with decreasing radius; in
case the heating sources (stellar mass loss; supernova heating) are
not strong enough to balance the cooling, a pressure-driven ``cooling
flow'' will occur, in which the gas flows subsonically inward,
where it converts into emission-line filaments, cooler gas and/or
(low-mass) stars. 
Typical quoted values for the mass inflow rates derived from the X-ray
data are $0.1 - 1$ \Mzon\ yr$^{-1}$ for giant elliptical galaxies
which are {\it not\/} at the centre of clusters (Thomas \etal
\cite{thom+86}), and up to several 10$^2$ \Mzon\ yr$^{-1}$ for central
dominant galaxies in clusters (e.g., Steward \etal \cite{stew+84}). 

An alternative concept is the ``evaporation flow'' scenario
[independently proposed by Sparks, Macchetto \& Golombek (\cite{spa+89})
and de Jong \etal (\cite{dej+90})] in 
which clouds of cold gas and dust have been accreted during
post-collapse galaxy interactions. In this scenario, thermal interaction
between the cool accreted gas and dust and the pre-existing hot gas
locally cools the hot gas (thus mimicking a cooling flow) while heating
the cool gas and dust, thereby giving rise to optical and far-infrared
emission. A prediction of this scenario is that the X-ray-emitting gas
distribution should follow that of the optical nebulosity quite
closely (Sparks \etal \cite{spa+89}; de Jong \etal \cite{dej+90};
Sparks \cite{spa92}), and that dust should be associated with the
optical nebulosity. 

In order to properly assess the appropriateness of either of these two
scenarios, it is important to study the relation between the
hot and cooler components of the ISM. To date, the study of this
relation has been restricted to statistical comparisons between the
total \Ha\ and X-ray luminosities of early-type galaxies detected by
the {\it Einstein\/} satellite (Shields \cite{shie91}; Trinchieri \&
di Serego Alighieri \cite{tdsa91} [hereinafter TdSA]; Goudfrooij
\cite{paul97}; Macchetto \etal
\cite{duccio+96}). These studies have 
produced somewhat ambiguous conclusions, which is most probably due to
different observational flux thresholds:\ Shields (\cite{shie91})
found essentially {\it no\/} correlation, whereas TdSA, Goudfrooij
(\cite{paul97}) and Macchetto \etal (\cite{duccio+96})
found that ---on 
average--- galaxies with a larger content of hot gas also have more
powerful line emission (often extended, at low surface
brightness). However, the scatter in this relation is significant
(like most of the correlations that involve X-ray emission),
and clearly needs to be better understood. 

In order to study the relation between the hot and cooler components
of the ISM in detail, high spatial resolution imaging at X-ray and
optical wavelengths of X-ray bright, \Ha-emitting early-type galaxies
are called for. Here we present the results of such a study for NGC~5846,
for which a detailed comparison between high spatial resolution X-ray data
and optical broad-band and narrow-band CCD imaging has been possible.

\section{NGC 5846}
\label{N5846}

NGC 5846 (= UGC~9706 = PGC~53932) is a giant E0 galaxy (de Vaucouleurs
et al.\ \cite{rc3}, 
hereinafter RC3) which dominates a small group of galaxies [group
LGG~398 of Garcia (\cite{garc93})], containing 8 members.  
NGC~5846 contains a compact radio core with a brightness of 20.9 mJy
at 20 cm (M\"ollenhoff, Hummel \& Bender \cite{moll+92}). {\it
Einstein\/} observations of NGC~5846 have revealed the presence of a
large reservoir of X-ray-emitting gas, extending out to 
$\sim\,120$ kpc (Biermann, Kronberg \& Schmutzler \cite{bier+89}),
which is equivalent to about 10 optical effective radii (cf.\ Bender,
Burstein \& Faber \cite{bbf92}). The total mass of hot gas is $\sim
10^{11} \Msun$, and the potential well of the hot gas corresponds to a
virial mass of almost $10^{13} \Msun$, which is very similar to that
of M\,87, the dominant elliptical of the Virgo cluster (Biermann \etal
\cite{bier+89}). Global galaxy properties of NGC~5846 are listed in
Table~\ref{tab5846}.  

\begin{table}
\caption[ ]{Global properties of NGC~5846.}
\label{tab5846}
\begin{tabular*}{8.75cm}{@{\extracolsep{\fill}}lll@{}} \hline \hline
\multi{3}{c}{~~} \\ [-1.8ex]                                                
 Parameter & Value & Reference \\ [0.5ex] \hline
\multi{3}{c}{~~} \\ [-1.8ex]  
RA~~(J2000.0)     & 15$\!^{\tinm{d}}$06$\!^{\tinm{m}}$29\fs17 & MHB92 \\
DEC~(J2000.0)     & +01\degr36$'$\,20\farcs9    	& MHB92 \\
   Galaxy Type    & E0        	    	   	& RC3 \\
                  & S0$_1$(0) 	     	   	& RSA \\
$r_{\scrm{eff}}$  & 69$''$			& BBF92 \\
$\sigma_0$        & 277 km s$^{-1}$ 		& D+87 \\
$B_T$             & 11.05 			& RC3 \\
$A_{B,\,\scrm{foreground}}$ & 0.14		& BH84 \\
$(B-V)_{\scrm{eff}}$ & 1.03 			& RC3 \\
$(U-B)_{\scrm{eff}}$ & 0.57 			& RC3 \\
$v_{\scrm{hel}}$  & 1822 km s$^{-1}$		& RC3 \\
$v_{\scrm{grp}}$  & 1595 km s$^{-1}$		& F+89 \\
Distance	  & 31.9 Mpc 			& {\tt *} \\
$M_{B_T}^0$       & \llap{$-$}21.99		& {\tt **} \\
\multi{3}{c}{~~} \\ [-1.8ex] \hline 
\multi{3}{c}{~~} \\ [-1.8ex] 
\end{tabular*}
 
\baselineskip=0.97\normalbaselineskip
{\small
\noindent 
{\sl Notes to Table~1.} \\
{\it Parameters:\/~}$r_{\scrm{eff}} \cor$ Effective radius;
$\sigma_0 \cor$ central velocity dispersion; $B_T \cor$ Total B
magnitude; $A_{B,\,\scrm{foreground}} \cor$ Absorption in B band
due to ISM in our Galaxy; $(B-V)_{\scrm{eff}}$,
$(U-B)_{\scrm{eff}} \cor\;$Mean colours inside the effective radius;
$v_{\scrm{hel}} \cor\;$Heliocentric velocity;
$v_{\scrm{grp}} \cor\;$Group velocity; $M_{B_T}^0 \cor\;$Total
absolute B magnitude, corrected for Galactic absorption. \\
{\it References:\/}~BBF92 $\cor$ Bender \etal (1992); 
BH84 $\cor$ Burstein \& Heiles (1984); 
D+87 $\cor$ Davies \etal (1987); 
F+89 $\cor$ Faber \etal (1989); MHB92 $\cor$ M\"ollenhoff \etal (1992); 
RC3 $\cor$ de Vaucouleurs \etal (1991); 
RSA $\cor$ Sandage \& Tammann (1987); {\tt *} $\cor$ derived from
$V_{\scrm{grp}}$; {\tt **} $\cor$ derived from $B_T$ and
$A_{B,\,\scrm{foreground}}$.
}

\end{table}

\section{Observations and data analysis}
\label{obs}

%
%
%
We obtained deep CCD images of NGC 5846 on May 28, 1996. We used the
3.5-m New Technology Telescope of the European Southern Observatory
(ESO) on Cerro La Silla, Chile. The red arm of the EMMI instrument was
used, which operates at f/5.2. The detector was a thin,
back-illuminated CCD of type Tektronix TK2048EB Grade 2, having
2048$\times$2048 sensitive pixels. We employed the ``fast'' readout 
mode, featuring a read-out noise of 5.2 e$^-$. The pixel size was
$24\mu\mbox{m}\!\times\!24\mu$m, yielding a scale of 0.27 arcsec per
pixel. NGC 5846 was observed through Bessel $V$ and $I$ filters, with
total exposure times of 27 and 18 min, respectively. Multiple exposures
were taken to enable easy removal of cosmic radiation events. The
seeing was 1\farcs0 in $V$ and 0\farcs8 in $I$, and the weather was
photometric. 

\subsection{CCD data reduction}

We used the ESO MIDAS image processing system for the basic reduction
of the CCD images. We co-added 20 ``bias'' frames to construct a master
bias frame which is free of cosmic radiation events and for which the
effect of read-out noise is negligible. This master bias frame was
subtracted from the science images after having been matched in
absolute level, using the serial overscan region of the CCD. The
images have been flat fielded using both dome flat fields illuminated by
scattered sunlight and twilight sky exposures. The dome flat fields
were used to correct for high spatial frequency variations, \eg
pixel-to-pixel variations. The twilight sky exposures were used to correct for 
the low-frequency spatial variations, since their colour temperature
is better matched to that of the science images obtained at night. 

Cosmic radiation events in the individual images were removed by an
appropriate averaging program. First, the images taken through the
same filter were aligned to a common coordinate system, using the
centroids of stars in the field of view. This alignment
procedure was accurate to 
within 0.03 pixel. The images were subsequently averaged together by
comparing all individual pixel values (per unit exposure time) with
the median value over all frames (taken using the same
filter). Individual pixels are rejected in the averaging process if
their value exceeds the range expected from the (sky + read-out) noise. 

\subsection{Surface photometry}
Ellipses were fitted to the isophotes of the galaxy images using the 
``ellipse'' fitting program in the  {\sf stsdas.analysis.isophote}
package, running within the {\sc iraf\/}\footnote{{\sc iraf\/} is
distributed by the National Optical Astronomy Observatories, which is
operated by the Association of Research in Astronomy, Inc., under
cooperative agreement with the National Science Foundation, U.S.A.}
image processing system.  
The position angle, ellipticity, and centre of the ellipses were 
free parameters in the fit, as long as the signal-to-noise was
sufficient. The residuals from the fit are parameterized in terms of
Fourier coefficients (see Jedrzejewski \cite{jedr87}). 
Before the model fits were executed, image pixels occupied by
foreground stars and neighbouring galaxies were flagged and ignored 
in the further analysis. Residual deviant sample points at each fitted
isophote were identified (and ignored) using a $\kappa$--sigma
clipping algorithm. 

The sky background level and its uncertainty was determined by fitting
power laws to the outer parts of the model intensity profiles as
described in Goudfrooij \etal (\cite{paul+94a}). 

Absolute flux calibration was achieved through observations of
standard stars from Graham (\cite{grah82}). Care was taken to choose
stars with colours similar to elliptical galaxies to eliminate the
dependence on colour terms in the calibration. The observations were 
corrected for atmospheric extinction using the values in the 1993
version of the ESO Handbook (Schwarz et al.), and for Galactic
foreground extinction using the $A_B$ value in Burstein \& Heiles
(\cite{burhei84}), listed in Table~\ref{tab5846}. We assumed $A_V = 0.75
A_B$ and $A_I = 0.45 A_B$. These extinction coefficients have been
calculated using  the standard Galactic extinction curve (e.g., Mathis
\cite{mathis90}) and the filter specifications given by Bessell 
(\cite{bess79}).  

After a successful fitting procedure, model images with purely
elliptical isophotes were reconstructed from the fit. Model
intensities of individual image pixels were reconstructed by a spline
fit to the radial intensity profile of the elliptical fit. Subtraction
of the model fit from the original image was done as a sanity check of
the quality of the fit. 

\section{Results}
\label{results}

\subsection{Surface brightness profile}
\label{sbprof}
\begin{figure}
\centerline{
\psfig{figure=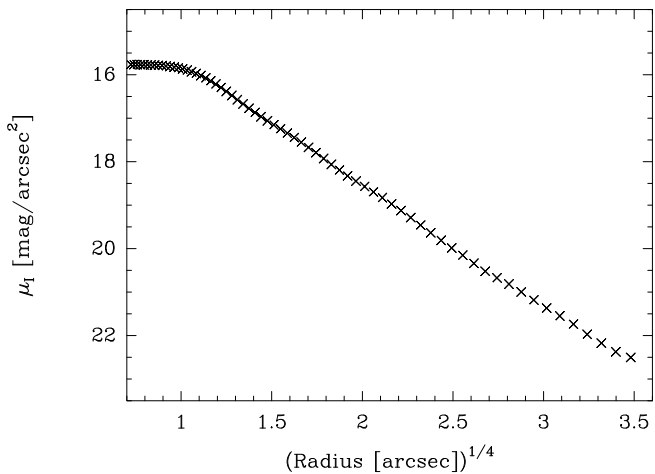,width=8.7cm}
}
\caption[]{$I$-band surface brightness versus radius$^{1/4}$ for NGC~5846}
\label{f:deV5846}
\end{figure}

\begin{figure*}
\centerline{
\psfig{figure=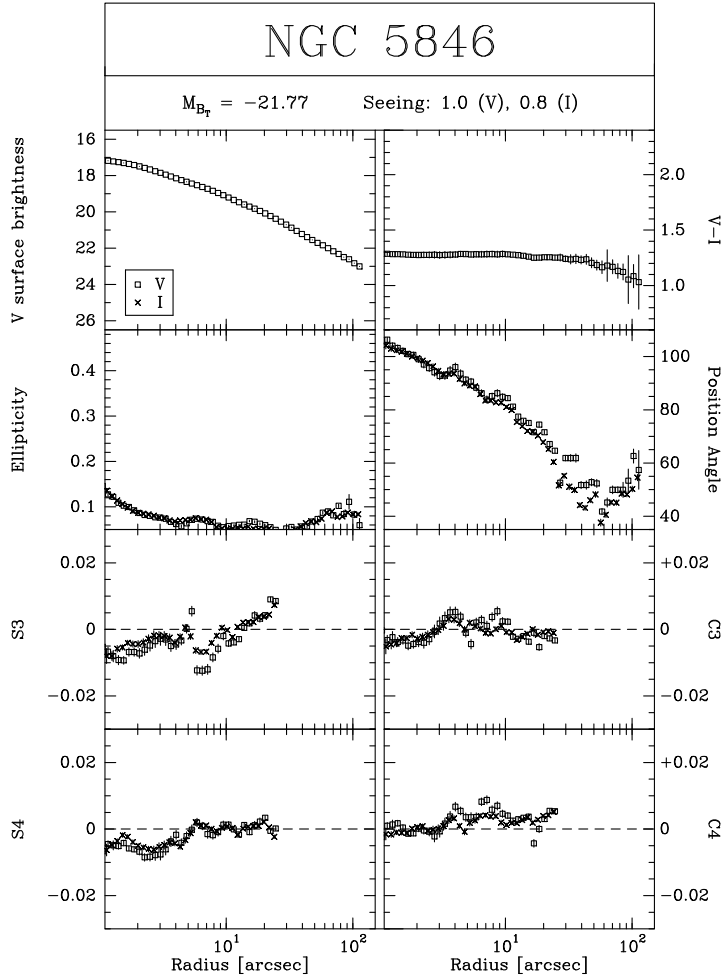,height=13cm}
}
\caption[]{The radial profiles of $V$ surface brightness, 
$V\!-\!I$ colour index, ellipticity, position angle, and the sine and
cosine 3$\theta$ and 4$\theta$ terms (where $\theta$ is the position
angle with respect to the major axis of the ellipse; these Fourier
terms are referred to as {\sf S3}, {\sf S4},  {\sf C3}, and {\sf C4})
for NGC~5846. The different symbols indicate data from different
passbands, as shown in the legend of the surface brightness plot}
\label{f:phot5846}
\end{figure*}

The results of the surface photometry 		
are given in graphical form in Figs.~\ref{f:deV5846} and
\ref{f:phot5846}. Outside the inner few arcsec where seeing limits
any further increase in surface brightness, the radial surface
brightness profile of NGC~5846 closely follows a de Vaucouleurs
(\cite{devau59}) law (cf.\ Fig.~\ref{f:deV5846}), confirming its RC3
classification as an E0 galaxy. The $V\!-\!I$ colour gradient is
very small:\ the colour stays essentially constant out to
$\sim$\,40$''$ (cf.\ Fig.~\ref{f:phot5846}).  

\subsection{High-order Fourier terms}
Fig.~\ref{f:phot5846} shows the presence of significant third-order
Fourier terms in NGC~5846, especially in the inner 10$''$, which are more 
significant in $V$ than in $I$. This behaviour is expected in the
presence of dust features (see Goudfrooij \etal \cite{paul+94a}). Also
the cos 4$\theta$ term seems to be influenced by dust. Taking into
account the influence of dust, the only significant deviation of the {\it
stellar\/} light from pure ellipses seems to be a slightly negative
sin 4$\theta$ component in the inner 5$''$, which indicates some
excess light at about 67\fdg5 from the major axis of the galaxy.  

\subsection{Dust in the central region}
\label{dustmeth}

Subtraction of an elliptical model from the galaxy image
reveals distinct, filamentary dust features in a ``boomerang'' shape,
reaching out to $\sim$\,12$''$ from the nucleus, beyond which they
become less obvious to detect. Tentative detections of dust were already
reported by TdSA and Buson \etal (\cite{buson+93}), who however did not
pursue any further quantitative investigation and did not have multi-colour
imagery. V\'eron-Cetty \& V\'eron (\cite{vv88}) did not find dust in
NGC~5846. 

\begin{figure}
\centerline{
\psfig{figure=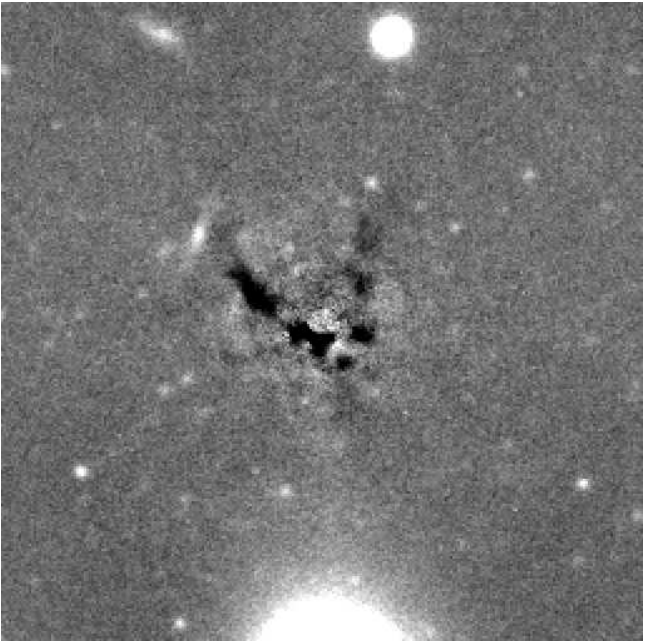,width=8.5cm}
}
\caption[]{Positive grey-scale reproduction of the central 70$''\times$70$''$
of NGC~5846. North is up and east is to the left. A model consisting
of purely elliptical isophotes has been subtracted from the $V$-band
image of the galaxy. Several dust features can be seen in the inner
regions, as well as globular clusters associated with NGC~5846 and
background disk galaxies. The bright stellar object 30$''$ north of
NGC~5846 is a foreground star, and the extended object at the lower
edge of the image is NGC~5846A, a dwarf elliptical galaxy 
}
\label{f:resv5846}
\end{figure}

Given the existence of the dust features, we reiterated the ellipse
fitting procedure once again, now with the dusty regions flagged and
discarded from the fit, so that the fitted ellipses accurately
describe the underlying stellar light distribution.  
A grey scale plot of the residuals of the final fit 
is shown in Fig.~\ref{f:resv5846}\footnote{Note,
however, that this method does not reliably reveal dust features
in the inner few seeing disks from the centre.}. A qualitative check on
the reality of the dust features in the central part ($r \leq 10''$)
---where the uncertainties in the ellipse fit are arguably largest--- is
provided by the fact that they are also clearly seen in the HST/WFPC2
image published by Forbes, Brodie \& Huchra (\cite{forb+97}) in their
study of the globular cluster system of NGC~5846.

\section{Properties of the dust in NGC~5846}
\label{dustprop}

\subsection{Extinction curve}
\label{extcurve}

To investigate quantitative properties of the dust extinction in
NGC~5846, we used the final model images to create ``extinction''
maps, 
\begin{equation}
A_{\lambda} = 
   -2.5 \log \left( \frac{I_{\lambda,\, \scrm{obs}}}{I_{\lambda, \, 
                    \scrm{model}}} \right)
\end{equation}
where $I_{\lambda,\, \scrm{obs}}$ is the observed
intensity level (of any individual image pixel), and $I_{\lambda,\,
\scrm{model}}$ is the intensity level of the purely elliptical model
image described above. The $A_I$ image was aligned with the $A_V$ image, and
rebinned to a common coordinate system. The uncertainty of the
alignment was 0.02 pixel. After slightly smoothing the (rebinned) $A_I$
image to match the seeing of the $V$ image, masks were set up covering
the regions occupied by dust. Numerical values for $A_V$ and $A_I$ were
then extracted within those masks for independent, rectangular boxes,
comparable in size to the seeing. The scatter within each box was
used (along with the sky background subtraction uncertainty) to
estimate the uncertainty associated with these measurements. 

\begin{figure}
\centerline{
\psfig{figure=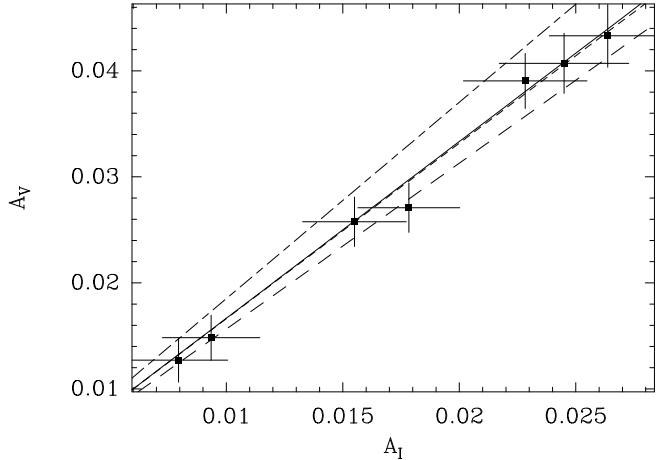,width=8.7cm}
}
\caption[]{$A_V$ versus $A_I$ for dust features in NGC~5846. The 
Galactic extinction curves from Mathis (1990) are drawn for
$R_V$ = 3.1 (solid line) and $R_V$ = 5.0 (long-dashed line). The
extinction curve for $R_V = 2.5$ from Steenman \& Th\'e (1989) is
also drawn (long-short-long dashed line). The short-dashed line
represents the extinction curve for $R_V$ = 3.1 in case 50\% of the
stars in NGC~5846 are in front of the dust for any given line of sight
}
\label{f:AI_AV_5846}
\end{figure}

A plot of $A_V$ versus $A_I$ of these boxes is shown in Fig.\
\ref{f:AI_AV_5846}, along with models for different extinction
curves. As Mathis (\cite{mathis90}) concluded in his review, the
Galactic interstellar extinction curve from the far-IR to the UV is
mainly a function of $R_V = A_V/E_{B-V}$. In Fig.\
\ref{f:AI_AV_5846} we have drawn the relation between $A_V$ and $A_I$
for $R_V$ = 3.1 (average value for dust in Galactic ISM) and 5.0 (dust
in outer clouds of our Galaxy). In addition, we have drawn the
extinction curve for $R_V$ = 2.5 as tabulated by Steenman \& Th\'e
(\cite{stethe89}). Low values for $R_V$ such as this ---which indicate
a relatively low value of the abundance ratio of large grains to
small grains--- are commonly found in elliptical galaxies with
large-scale dust lanes, which is thought to be the result of evolved
destruction of dust grains since the time the dust in the rapidly
rotating dust lanes was accreted from outside 
(Goudfrooij \etal \cite{paul+94c}). 

Figure~\ref{f:AI_AV_5846} shows that the extinction values for the
dust in NGC~5846 are best fit with the curve for $R_V$ = 3.1, the
canonical value for dust in our Galaxy. While these data cannot really
dismiss other values of $R_V$ (especially values slightly larger than
3.1), this means that the dust in NGC~5846 is consistent with having
``normal'' Galactic properties. This result holds when taking into account the 
effect of the presence of foreground stars (see, e.g.,
N{\o}rgaard-Nielsen \etal \cite{hunn+93}) to the appearance of the
extinction curve (cf.\ Fig.\ \ref{f:AI_AV_5846}). 
In the scenario of Goudfrooij \etal (\cite{paul+94c}), this means that
the dust in NGC~5846 is relatively young (younger than $\sim\,$10$^9$
yr), which is consistent with the filamentary structure of the dust lane. 

\subsection{Mass of dust}
\label{m_d}

The amount of dust can be derived from the $A_{\lambda}$ values by
assuming a grain size distribution as well as a composition of the
dust grains (see, e.g., Goudfrooij \etal \cite{paul+94c}). For a given
grain size distribution function $n(a)$ where $a$ is the grain radius,
the cross-section for spherical particles at  wavelength $\lambda$ can
be written as 
\begin{equation}
C_{\scrm{ext}}(\lambda) = 
  \int_{a_{-}}^{a_{+}} Q_{\scrm{ext}}(a,\lambda)\; \pi a^2 \, n(a)
   \; {\mbox{d}}a 
\end{equation}
(see, \eg Spitzer \cite{spit78}), where $a_{-}$ and $a_{+}$ represent the 
lower and upper cutoffs of the 
grain size distribution, respectively, and $Q_{\scrm{ext}}(a,\lambda)$ is
the extinction efficiency at wavelength $\lambda$. The total
extinction in magnitudes at wavelength $\lambda$ can then be written as 
\begin{equation}
A_{\lambda} = 1.086 \: C_{\scrm{ext}}(\lambda) \times l_{\scrm{d}}
\end{equation}
where $l_{\scrm{d}}$ is the length of the dust column in the line of sight. 
The dust column density $\Sigma_{\scrm{d}}$ [g cm$^{-2}$] is 
\begin{equation} 
\Sigma_{\scrm{d}} = 
   \int_{a_{-}}^{a_{+}} \frac{4}{3} \pi a^3 \, \rho_{\scrm{d}} \; n(a)
   \; {\mbox{d}}a \times l_{\scrm{d}} 
\label{dustcoldens}
\end{equation} 
where $\rho_{\scrm{d}}$ is the specific grain mass density, which we
assume to be 3 g cm$^{-3}$ (Draine \& Lee \cite{dralee84}; this
represents the average of their graphite and silicate grain models). Hence,
dust masses can be estimated from the measured \Av\ values by making
sensible assumptions for $a_{-}$, $a_{+}$, $n(a)$ and
$Q_{\scrm{ext}}(a,V)$.  
 
From the available data, we have reason to believe that the dust in
NGC~5846 is similar to that in our Galaxy (cf.\ Sect.\
\ref{extcurve}). Hence, we assume the grain size distribution to be
the Galactic one ($n(a) \propto a^{-3.5}$, cf.\ Mathis, Rumpl \&
Nordsieck \cite{mrn77})\footnote{Note that we only have information
about the ``large'' grains (radius of order 0.1 $\mu$m) which absorb optical
light. Our measurements are not sensitive to extinction by small
grains, which has to be studied in the UV.}. Using extinction
efficiencies for graphite and ``dirty silicate'' grains listed by
Goudfrooij \etal (\cite{paul+94c}), and assumed equal abundances of 
graphite and silicate grains (cf.\ Draine \& Lee \cite{dralee84}), the
resulting dust mass is $M_{\scrm{d}} \sim 7.1\;10^3\; D_{31.9}^2$
\Mzon, where $D_{31.9}$ is the distance in units of 31.9 Mpc. Here we
assumed that 50\% of the stars in NGC~5846 are in front of the dust
for any given line of sight. 

\subsection{Temperature of dust; far-infrared emission}
\label{t_d}

\begin{figure}
\centerline{
\psfig{figure=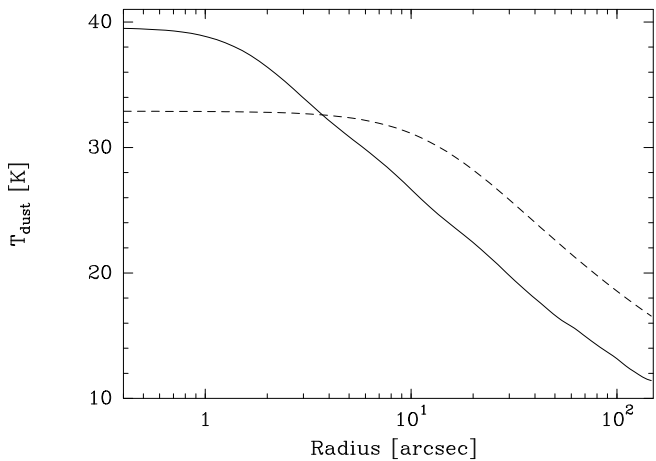,width=8.7cm}
}
\caption[]{Plots of the local dust temperature due to stellar heating
(solid line) and due to heating by hot electrons in the X-ray-emitting
gas (dashed line) versus distance from the centre of NGC~5846}
\label{f:t_d_5846}
\end{figure}

Following the methods detailed upon by Goudfrooij \& de Jong
(\cite{gdj95}), we have calculated the expected radial dust temperature
profile for NGC~5846. 
Being embedded in a luminous X-ray-emitting halo, the dust grains in
NGC~5846 are heated by two main processes:\ {\it 
(i)\/} stellar heating (the ``general interstellar radiation field''),
and {\it (ii)\/} collisions with hot electrons in the X-ray-emitting
gas (Dwek \cite{dwek86}; de Jong \etal \cite{dej+90}). 
We used the surface brightness profiles presented in Sect.\
\ref{sbprof} to derive the radial gradient of the average intensity of
optical radiation. 
As to the heating by hot electrons, we used the radial
electron density profile of Trinchieri \etal (\cite{trin+97}),
which was derived from ROSAT HRI images. For the 
electron temperature we assumed $kT_{\scrm{e}} = 0.56$ keV which is the
average temperature in the $0-1$ arcmin range (Trinchieri \etal
\cite{trin+97}). 
Dust temperatures were then calculated by equating the heating rates
with the cooling rate of a dust grain by infrared emission (see
Goudfrooij \& de Jong \cite{gdj95} for further details). The resulting 
radial dust temperature profiles are presented in Fig.~\ref{f:t_d_5846}.
The dust temperatures were determined for a grain size of $a$ = 0.1 $\mu$m, and
therefore representative of the ``large'' grains, which absorb optical
light. Inclusion of the small grains with radius $a \la 0.01\;\mu$m, 
which are stochastically heated to high temperatures (e.g., Dwek
\cite{dwek86}), can in principle raise the average grain temperature
somewhat, especially in a case where the grain size distribution
includes a relatively substantial amount of small grains. However, we
believe that this is not the case in X-ray bright ellipticals such as
NGC~5846, for the following reasons: {\it (i)\/} the sputtering time
for small grains in a hot 
gas with $T_e \sim 10^7$ K is extremely short, e.g., $\la 10^6$ yr for
grains of radius 0.01 $\mu$m (cf.\ Sect.~\ref{dustorig}), so that they
will be preferentially removed from the grain size distribution; {\it
(ii)\/} the 12 and 25 $\mu$m IRAS flux densities of giant elliptical
galaxies rule out any significant contribution from emission by small
grains as they are entirely consistent with the combination of the
contributions of stellar (photospheric) and circumstellar dust around
late-type stars (cf.\ Knapp et al.\ \cite{knapp+92}; Goudfrooij
\cite{paul94t}).  
Furthermore, the appropriate average value for a grain radius when
determining the average grain temperature is that weighted by the
contribution of each grain to the infrared emission and hence by the grain
volume $a^3$: 
\[ \lmean a \rmean \, = \, \frac{\int_{a_{-}}^{a_{+}} a\, n(a)\, a^3
\; \mbox{d}a}{\int_{a_{-}}^{a_{+}} n(a)\, a^3 \; \mbox{d}a} \]
(cf.\ also Hildebrand \cite{hild83}). For a grain size distribution
such as the one proposed by Mathis et al.\ (\cite{mrn77}; $n(a)
\propto a^{-3.5}$), we obtain $\lmean a \rmean$ = 0.1 $\mu$m, the
grain size we used to determine the grain temperature. 

We have checked whether our derived dust temperature distribution is
consistent with the {\sl IRAS\/} data on NGC~5846 (Knapp \etal
\cite{knapp+89}). Assuming a $\lambda^{-1}$ emissivity law at
wavelengths $\lambda \la 100 \mu$m, the {\sl IRAS\/} emission from
(Galactic) dust is  
\begin{equation}
S_{\nu}  =  1.96 \times 10^{10}\; D_{\scrm{Mpc}}^{-2}\: \lambda_{\mu}^{-4}\; 
 \frac{M_{\scrm{d}}}{\exp(1.44\:10^{4}/\lambda_{\mu}\,T_{\scrm{d}}) - 1}  
\label{eq:irasflux}
\end{equation} 
(cf.\ Goudfrooij \& de Jong \cite{gdj95}) 
where $S_{\nu}$ is the {\sl IRAS\/} flux density in mJy,
$D_{\scrm{Mpc}}$ is the distance of the Galaxy in Mpc, $M_{\scrm{d}}$
is the dust mass in solar units, $\lambda_{\mu}$ is the
wavelength in $\mu$m, and $T_{\scrm{d}}$ is the dust temperature. 
We extracted the $A_V$ values of image pixels occupied by dust
features. Dust temperatures, and hence {\sl IRAS\/} flux densities at
60 and 100 $\mu$m (cf.\ Eq.\ \ref{eq:irasflux}) were then calculated
for each image pixel, according to their (projected) distance from the
galaxy nucleus. The resulting total flux densities are $S_{60} = 12.1$
mJy 				
and $S_{100} = 22.2$ mJy, 	
which is consistent with the 1\,$\sigma$ upper limits from {\sl IRAS}:
$S_{60} < 37$ mJy and $S_{100} < 112$ mJy (Knapp \etal \cite{knapp+89}).  


\section{Comparison between different components of the ISM in NGC~5846}
\label{compar} 

\begin{figure}
\begin{minipage}[t]{8.8cm}
\centering
\ \\ [-1ex]
\centerline{\psfig{figure=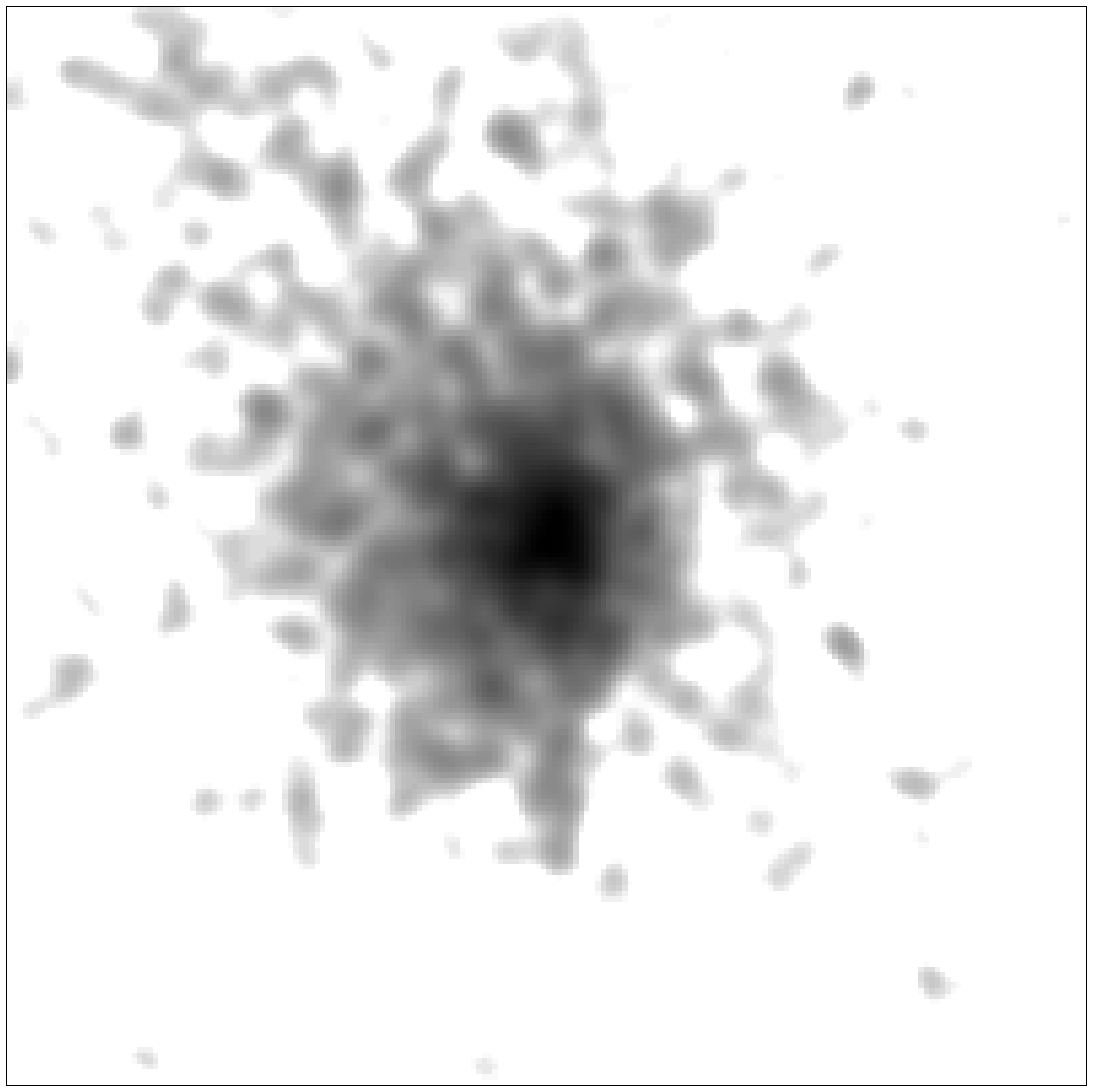,height=6.7cm}}
\vspace*{-0.7mm}
\centerline{\psfig{figure=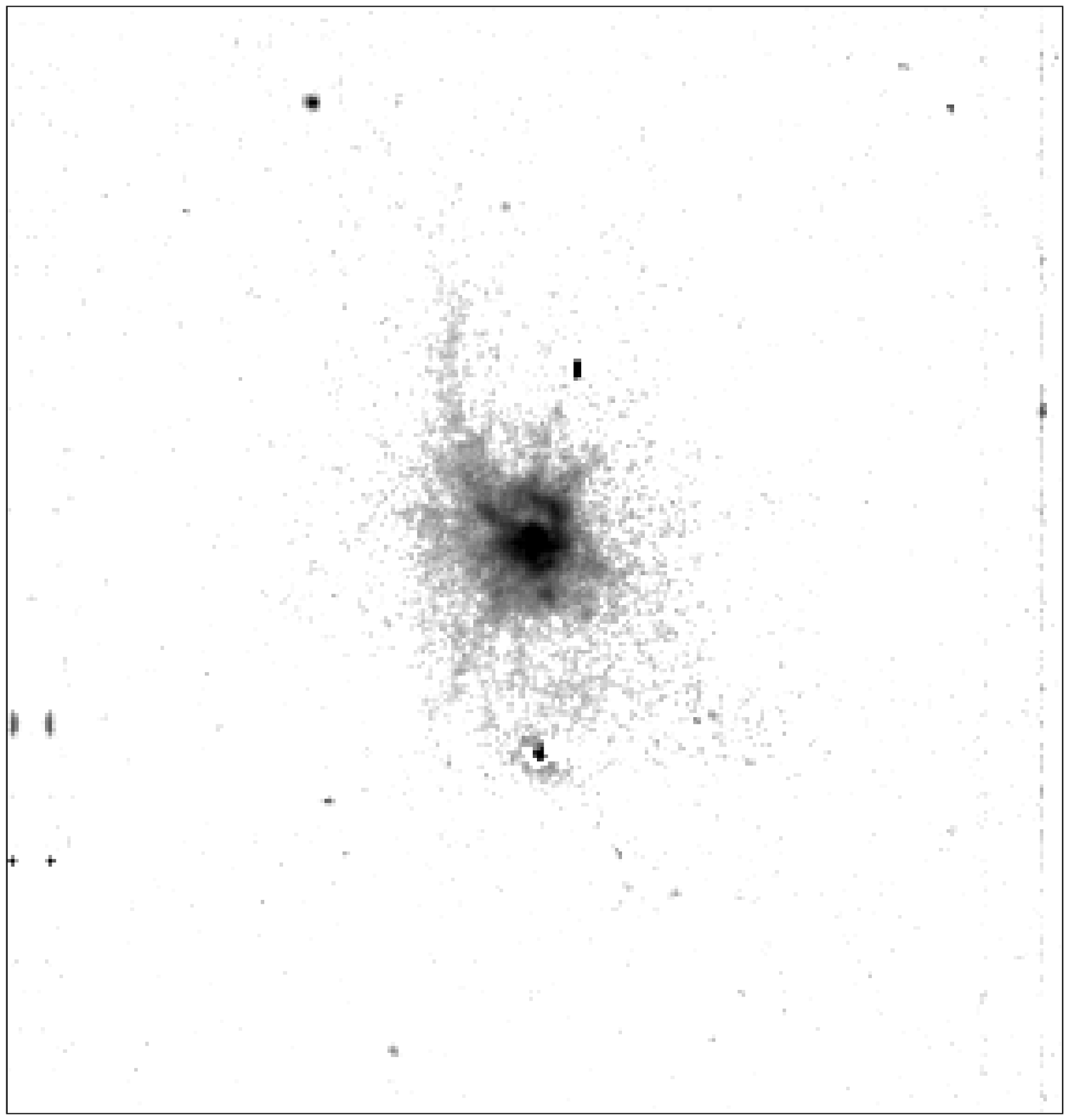,height=6.7cm}}
\vspace*{2mm}
\centerline{\psfig{figure=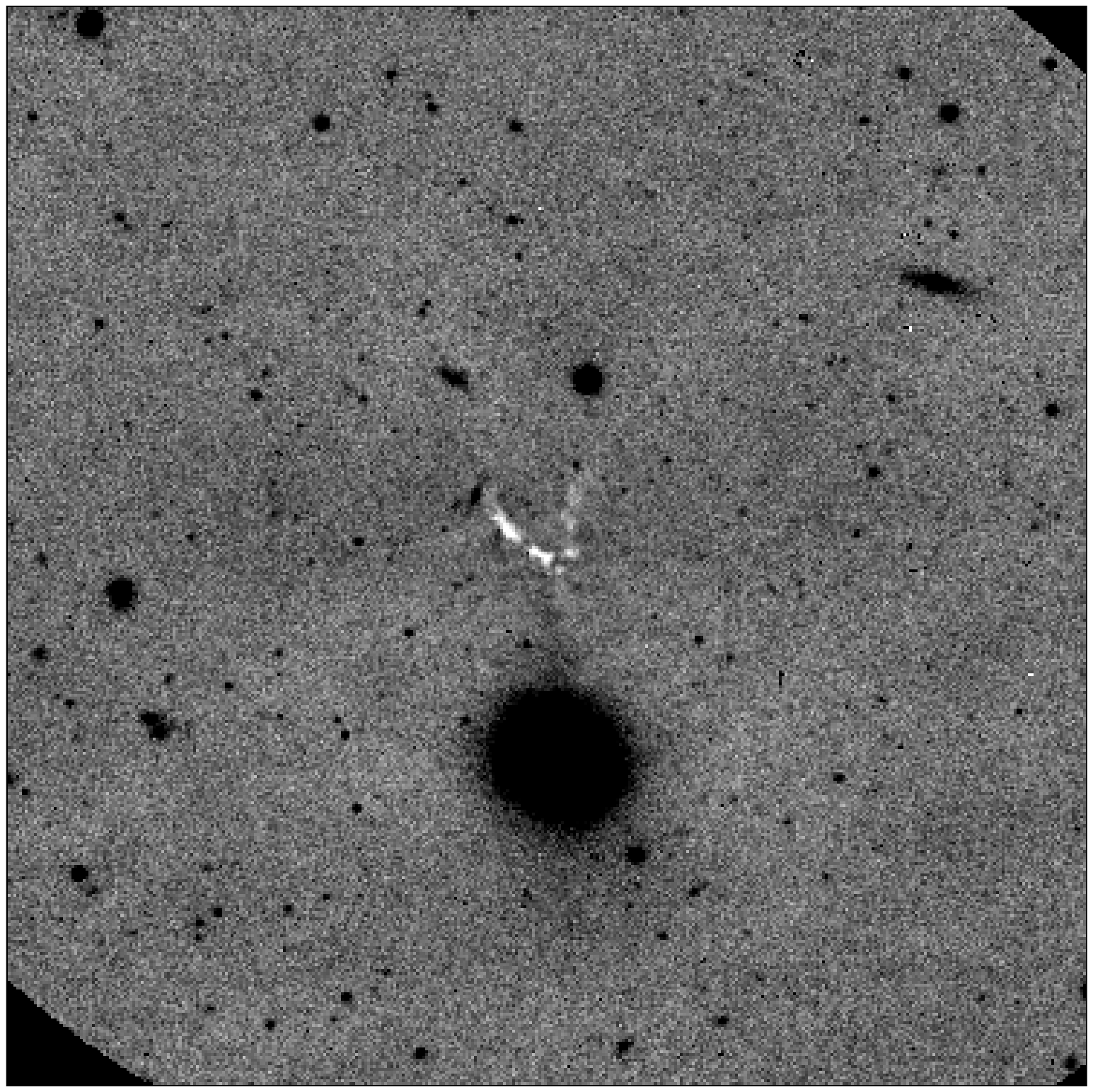,height=6.7cm}}
\end{minipage}
\caption[]{Grey-scale reproductions of the distributions of X-ray
emission (top),  \Ha+\NII\ emission (middle), and $A_V$ of dust
extinction in the central $200\times200$ arcsec of NGC~5846. North is
up and east is to the left. We used a
{\it logarithmic\/} grey-scale transfer table for the X-ray-- 
and \Ha+\NII\ images to enhance the low surface brightness
features. The X-ray map is the ROSAT HRI image from Trinchieri \etal
(1997), and the \Ha+\NII\ image was published before in TdSA
}
\label{f:x_ha_av}
\end{figure}

The distributions of dust, optical nebulosity, and X-ray
emission in the central 200\,$\times$\,200 arcsec of NGC~5846 are depicted
in Fig.~\ref{f:x_ha_av}. The ROSAT 
HRI map is from Trinchieri \etal (\cite{trin+97}), and the \Ha+\NII\ map 
is from TdSA.  
To suppress noise from the sky background, both of these two maps have been
slightly smoothed by convolution with circular Gaussians. The values
of $\sigma$ were 2\farcs5 and 0\farcs3, respectively. 
The X-ray-- and \Ha+\NII\ morphologies have been described in detail
before; here we only mention features that are relevant to the
discussion in Sect.~\ref{dustorig} and \ref{disc}. 

\subsection{Optical nebulosity vs.\ X-ray emission}

The morphology of the X-ray emission in NGC~5846 is highly asymmetric
and clumpy relative to that of the stellar (optical) emission
(NGC~5846 is a very round E0 galaxy). The asymmetry is mostly caused
by enhanced emission in the NE quadrant; interestingly, X-ray spectral
data from the ROSAT PSPC show that the emission in the NE quadrant is 
cooler than in the other three quadrants at the same
distance from the centre (Trinchieri \etal \cite{trin+97}). 
The distribution of \Ha+\NII\ emission in NGC~5846 is also rather
complex and asymmetric. Filamentary structures are visible at
different scales. The most extended filament is pointed towards the
NE, where the excess X-ray emission is also situated. 
All in all, there seems to be a spatial correlation
between X-ray and \Ha+\NII\ emission features, strongly suggesting a
close physical link between the ``hot'' and ``warm'' components of the
ISM.  

We note that this spatial correlation has also been found in NGC~1553
(Trinchieri \etal \cite{trin+97}) and NGC~4696 (Sparks, Jedrzejewski
\& Macchetto  \cite{spa+94}). 
All of these three X-ray-emitting ellipticals exhibit
X-ray morphologies that are much more clumpy and asymmetric than their
smooth optical (broad-band) isophotes. 

\subsection{Optical nebulosity vs.\ dust absorption}

\begin{figure}
\begin{minipage}[t]{8.8cm}
\centering
\ \\ [-1ex]
\psfig{figure=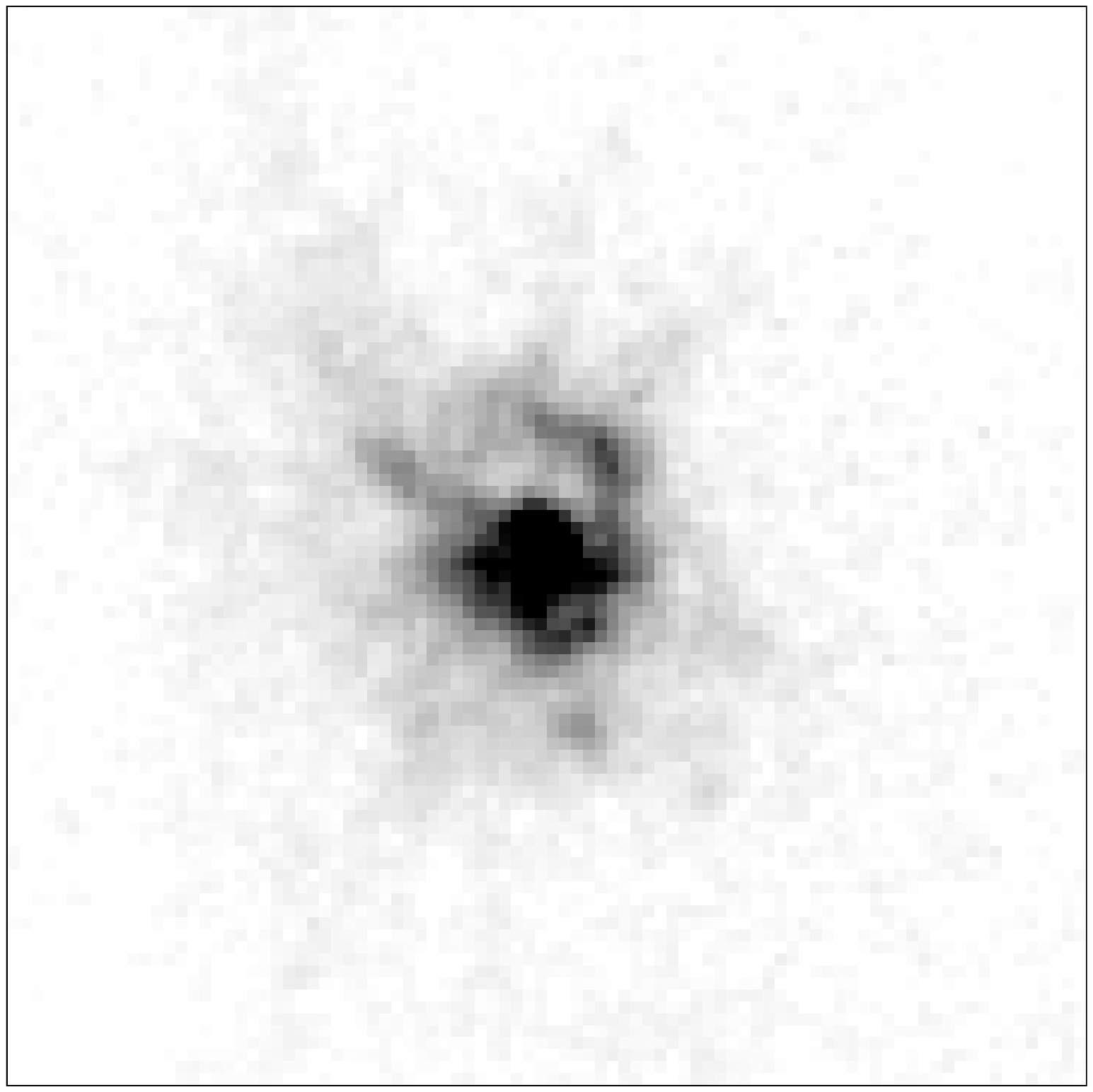,width=8.75cm} 
\vspace*{1mm}
\psfig{figure=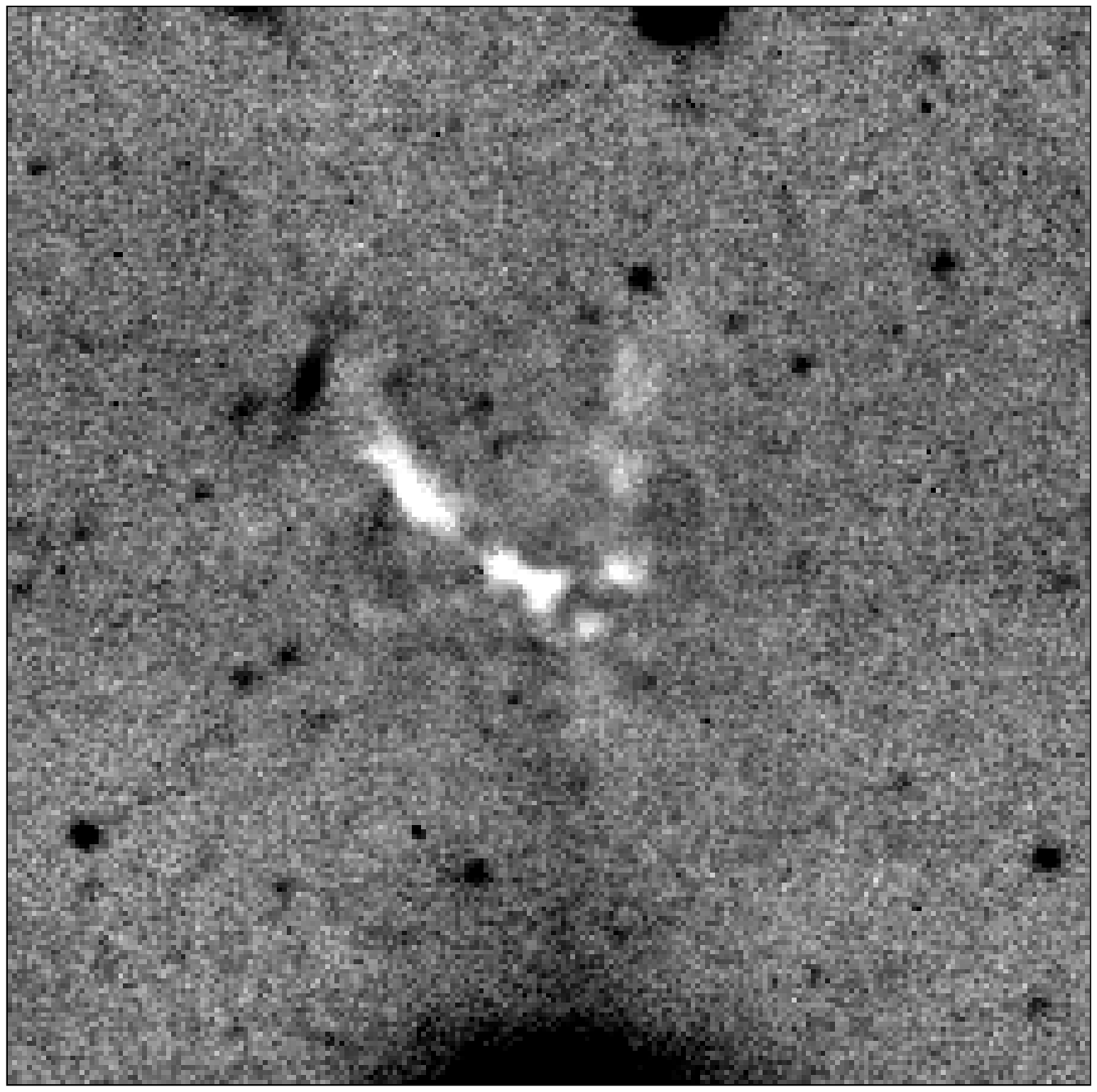,width=8.75cm} 
\end{minipage}
\caption[]{{\sf Top:} Grey-scale reproduction of the distribution of
\Ha+\NII\ emission in the central $60\times60$ arcsec of
NGC~5846. {\sf Bottom:} Grey-scale reproduction of the distribution of
$A_V$ of dust extinction in the central $60\times60$ arcsec of
NGC~5846. Maximum $A_V$ = 0.065 (white), and the faintest detectable dust
features have an $A_V$ of 0.015. 
A {\it linear\/} grey-scale transfer table 
was used for both images to show the similarity of the high surface
brightness features of dust and nebulosity
}
\label{f:ha_av}
\end{figure}

As the dust lane is small compared with the angular dimension of the
X-ray emission, we depict the distributions of dust and optical nebulosity
in the inner 60 $\times$ 60 arcsec of NGC~5846 separately in
Fig.~\ref{f:ha_av}. An important result of this comparison is that the
dust lane is associated with the filament system that has the highest
\Ha+\NII\ surface brightness in NGC~5846. Hints of the 
other, roughly radially distributed \Ha+\NII-emitting filaments (which
have lower surface brightness) are visible in the $A_V$ image when
inspected on a computer screen (and in the HST/WFPC2 image of Forbes \etal
\cite{forb+97}), but unfortunately hardly recognizable on the
grey-scale reproduction in Fig.~\ref{f:ha_av}. The remaining \Ha+\NII\
emission, which has a significantly lower surface brightness and a 
more spherically symmetric distribution, does not have an obvious
counterpart in the $A_V$ 
image. This does {\it not\/} imply that the nebulosity at low surface
brightness is not associated with dust, since the method used to
reveal the dust features (cf.\ Sect.~\ref{dustmeth}) is not
appropriate for detecting a diffuse distribution of dust. 

The finding of dust being associated with optical nebulosity in X-ray
bright elliptical galaxies is by no means limited to the case of
NGC~5846: this association has now been identified in several X-ray
bright ellipticals (cf.\ J{\o}rgensen \etal \cite{jor+83}; Hansen
\etal \cite{han+85}; Sparks \etal \cite{spa+89}, \cite{spa+93}; de
Jong \etal \cite{dej+90}; Goudfrooij \etal \cite{paul+94b}; Pinkney
\etal \cite{pink+96}). A physical connection between the dust and the
optical nebulosity seems to be required to explain this observed
association.  

\section{Origin of dust and ionized gas in NGC 5846}
\label{dustorig}

Taken at face value, the mere presence of dust features in the central
region of an X-ray bright
elliptical galaxy such as NGC~5846 is an important finding in the 
context of assessing the appropriateness of the ``cooling flow''
and/or the ``evaporation flow'' scenarios mentioned in Sect.\
\ref{intro}, since the local physical conditions are very hostile for
dust grains. In the following discussion, we will address implications
of the presence of dust in X-ray bright ellipticals to the
interpretation of properties of their multi-phase ISM, both
specifically for the case of NGC~5846 and in a more general
sense. 

\subsection{Cooling flow origin\,?}

The lifetime of a dust grain of radius $a$ against collisions with hot
protons and $\alpha$-particles (``sputtering'') in a hot gas with
$T_{\scrm{e}} \sim 10^7$ K is 
\begin{equation}
\tau_{\scrm{d}} \equiv a \left| \frac{\mbox{d}a}{\mbox{d}t} \right|^{-1}
   \simeq \; 2 \times 10^5 
   \left(\frac{n_{\tinm{H}}}{\mbox{cm}^{-3}}\right)^{-1}  
   \left(\frac{a}{0.1\,\mu\mbox{m}}\right)\; \mbox{yr}
\label{eq:tau_d}
\end{equation} 
(Draine \&\ Salpeter \cite{drasal79}; Tielens \etal \cite{tiel+94}),
which is typically of order only $\sim\,$10$^7$ yr for grains of
radius 0.1 $\mu$m in the central few kpc of X-ray bright ellipticals such as
NGC~5846 (and proportionally shorter for smaller grains). Hence, any
matter that condenses out of a ``cooling flow'' in the central regions of
early-type galaxies is very likely to be devoid of dust. This is 
illustrated by e.g., the finding that the intergalactic  medium within the
Coma cluster is depleted in dust by a factor of $\sim$\,140 with
respect to the Galactic dust-to-gas ratio (Dwek \etal \cite{dwek+90}).

The cooling gas is also unlikely to generate
dust internally: While pressures in the central regions of a cooling
flow ($nT \sim 10^5 - 10^6$ cm$^{-3}$\,K) are high compared
with typical pressures in the diffuse ISM in our Galaxy, they are
still significantly lower than those of known sites of grain formation
such as the atmospheres of red giant stars ($nT \sim 10^{11}$
cm$^{-3}$\,K; Tielens \cite{tiel90}).  

\subsection{Stellar mass loss\,?}

Dust and gas are continuously injected into the ISM of early-type
galaxies by red giant winds and planetary nebulae at a rate 
\begin{equation}
\alpha_n (t)\, \rho_{*} (r)\mbox{,}
\label{eq:injec}
\end{equation}
where $\rho_{*}$ is the stellar density (typically of order
10$^{-21}$ g cm$^{-3}$ in the galaxy centre), and $\alpha_n (t) \equiv
\dot{M_*}/M_* (t)$ is the specific rate of stellar mass loss ($\approx 5
\times 10^{-20}$ s$^{-1}$, essentially insensitive to the
stellar IMF in the case where the bulk of the stars were formed in a
coeval instantaneous burst, cf.\ Mathews \cite{math89}). The evolution
of internally produced dust in a hot, X-ray-emitting environment has
been studied by Tsai \& Mathews (\cite{tsamat95}, \cite{tsamat96}) who
found that the vast majority of dust grains at any galactocentric
radius have been introduced at that same radius, due to the steep
radial gradients of stellar and gas density and thus dust
lifetime. The gas and dust ejected from stars are expected to merge
essentially instantaneously with the hot gas (on time scales
$\sim$\,10$^4 - 10^5$ yr, Mathews \cite{math90}), where it is subject
to rapid erosion by sputtering.  

The continuous dust injection rate of Eq.\
(\ref{eq:injec}) amounts to $\sim 2 \times 10^{-3}$ \Mzon\
yr$^{-1}$ for the central 15$''$ of NGC~5846 ($\sim$\,2 kpc;
$B_{<15''}$ = 13.16). In a Hubble time, one would
then accumulate $\sim$\,10$^4$ \Mzon\ of internally produced
dust, taking into account the short sputtering time scale for grains in
that region (cf.\ Tsai \& Mathews \cite{tsamat96}). Although this is
equal (to within the uncertainties) to the mass in the dust features
(cf.\ Sect.\ \ref{m_d}), we have strong doubts that the observed dust
features are of internal origin. If that were the case, it is easy to
show that dust features similar to these would be easily observed
in the central kpc of {\it each and every\/} nearby luminous
elliptical. Although dust has been optically detected in a large
fraction of giant ellipticals, the mass of such dust certainly does
not correlate significantly with global properties of the galaxies
(Goudfrooij \etal \cite{paul+94b}; van Dokkum \& Franx 
\cite{vdf95}). It seems more probable that the gas and dust injected
by evolved stars into the hot gas generally ends up in a more diffuse
distribution, thus not easily detectable by optical means. Brighenti
\& Mathews (\cite{brimat96}) studied the evolution of internally
produced gas as a function of total angular momentum of the stellar
system, and found that when a small but typical amount of galactic
rotation is introduced, this gas will be deposited in a large
disk extending out to an effective radius or beyond. 


\subsection{Accreted during galaxy interaction\,?}
A simple explanation for the asymmetric 
morphology of the main dust feature, which accounts for the presence of
both dust and gas (as well as their physical association), is that
the dust and gas represent the debris of 
material that has been accreted or tidally stripped from a small
neighbouring galaxy. 
Comparison of the observed filamentary but
essentially coherent structure with recent model calculations of the
formation of dust lanes or rings in galaxies (e.g., Steiman-Cameron \&
Durisen \cite{stedur88}, \cite{stedur90}; Christodoulou \& Tohline
\cite{chrtoh91}) shows that the accretion event must have occurred recently,
$\la 0.2 - 2 \, \times 10^8$ yr ago, depending on the orientation of
the in-falling galaxy with respect to the equatorial plane of the giant
elliptical. In this respect, it is tempting to identify the extended
\Ha+\NII-emitting filament (pointing toward the NE) with a tidal feature
which might reflect the original direction from which the small galaxy
was accreted. Kinematic information from deep \Ha+\NII\ spectroscopy
would obviously shed more light on the nature of this filament.  

\subsubsection{Constraints on the nature of the donor galaxy}
\label{donor}

If a galaxy interaction has indeed taken place, the apparent absence of any
obvious ``stellar'' signs of a recent interaction such as ``shells''
or a secondary nucleus in NGC~5846 has to be accounted for. 

Shells are quite common among elliptical galaxies, and believed to be
stellar remnants of galaxy encounters (see e.g., Prieur
\cite{prie90}; Forbes \etal \cite{forb+94}, and references
therein). Many different model simulations have been put forward to
account for the presence  and morphology of shells (e.g., major
mergers, minor mergers, and weak interactions); however, the
simulations typically only covered a small range of parameter space as
to the initial conditions and orbital parameters for the
interaction. For instance, it remains to be seen how sensitive the
formation and evolution of shells is to the nature of the secondary
galaxy (e.g., mass ratios, galaxy types). The simulations to date
involved strongly nucleated secondary galaxies (bulges, small
ellipticals), whereas in-falling irregular galaxies (of Magellanic
type) may be expected to disrupt significantly faster (due to dynamical
friction) than strongly nucleated galaxies do, which would result in much
lower stellar densities in any shells that might still form. Another
(rather remote) possibility for not finding shells might be that most of the
stars from the secondary galaxy have not yet had time to form shells
(i.e., less than a crossing time), but in that case one would expect
to see the nucleus of the secondary galaxy (if it was indeed
nucleated). All in all, the absence of shells seems to favour a
scenario in which the secondary galaxy was {\em not\/} significantly
nucleated, e.g., a small irregular galaxy.  

The small amount of filamentary dust observed ($\la 10^4 \Msun$, 
cf.\ Sect. \ref{m_d}) also indicates that the secondary
galaxy must have been a small satellite. A galaxy similar to the Small
Magellanic Cloud ($M_{\scrm{dust}} = (1.4\pm0.2) \times 10^4 \Msun$,
cf.\ Schwering 1988, Chap.~6) seems to be a good candidate. 

\section{Discussion}
\label{disc}

\subsection{Survival of the dust}
\label{dustsurv}

Let us assume for the moment that the dust lane and ionized gas
represent remnants of a small, relatively gas-rich galaxy, that has recently
been accreted by NGC~5846. Adopting the velocity dispersion
of the NGC~5846 group of galaxies of $\sim$\,150 \kms\ (Garcia 1993)
as a typical asymptotic infall velocity, we estimate that an in-falling
galaxy crosses the X-ray core of 4.6 kpc diameter in about $3
\times 10^7$ yr. This provides a lower limit for the life time of the
dust in that region unless some form of replenishment is
taking place. 

How does this compare to the sputtering time scale for dust grains?
From the analysis of the ROSAT observations (Trinchieri \etal
\cite{trin+97}), the electron density of the hot gas in the inner
15$''$ of NGC~5846 (where the dust is observed) is $n_e = 0.03$ cm$^{-3}$.  
From Eq.\ (\ref{eq:tau_d}) we obtain a lifetime of $8 \times 10^6$ yr
for grains of radius 0.1 $\mu$m ($n_H = 0.83\, n_e$ for gas with a cosmic
abundance of Hydrogen and Helium). 
In principle, this time scale is not very much shorter than the
crossing time mentioned above, which seems to allow for the
possibility that the donor was a small irregular galaxy which
originally hosted a few times more dust than that observed at present, of
which part of the dust has sputtered away in the mean time. On the
other hand, it is at least as likely that (part of) the dust has
already spent at least a few crossing times around the nucleus of
NGC~5846, as evidenced by the multiple filaments of dust and nebulosity. In
that case, the remaining dust and gas clouds must have been
replenished to maintain the observed dust mass, since the in-falling
galaxy cannot have been a large one (cf.\ Sect.\ \ref{donor}). 
Note that this latter scenario would be even more likely in case any
significant number of small grains would still be present. 

\subsection{Maintaining the observed amount of dust: Evaporation off
cool gas clouds}
\label{evapflow}

\subsubsection{Cooling vs.\ evaporation}

For the observed dust mass of $7 \times 10^3$ \Mzon, the required dust
replenishment rate is $9 \times 10^{-4} \Msun$ yr$^{-1}$. Our working
hypothesis is that this replenishment takes place by evaporation off
cool gas clouds brought in during the galaxy interaction, as proposed
by de Jong \etal (\cite{dej+90}). 
Allowing a reasonable range for the gas-to-dust mass ratio of $100 -
800$ (the latter being appropriate to the Small Magellanic Cloud, cf.\
Bouchet \etal \cite{bouc+85}), the dust replenishment rate is equivalent
to a gas evaporation rate $\dot{M} \sim 0.1 - 0.7$ \Mzon\ yr$^{-1}$. 

How does this compare with the mass deposition rate derived from
the X-ray data? In a steady cooling flow, the mass-loss rate required
to support the flow is 
\[ \dot{M}(r) \la \int_{0}^{r} \frac{L_X(r')}{H(r')} \;\;\mbox{d}r' \]
(e.g., Thomas \etal \cite{thom+86}) where $L_X(r)$ is the X-ray
luminosity and $H(r) = 5\,kT_e(r)/2 \,\mu m_H$ is the enthalpy of
the gas (where $\mu m_H = 1.4 \, m_H$ is the atomic
mass per Hydrogen atom for a gas with cosmic Helium
abundance). Using the radial distributions of X-ray flux and
electron temperature 
of Trinchieri et al.\ (1997), we obtain $\dot{M}
\la 0.36$ \Mzon\ yr$^{-1}$ for NGC~5846, which is entirely consistent
with the gas evaporation rate found above. 

\subsubsection{Properties of evaporating gas clouds}

Cold gas clouds embedded in a hot gas are heated by fast-moving
electrons. Assuming that the clouds are much smaller than the mean
free path of hot electrons (``saturated heat flow''), the heat flow is
(Cowie \& McKee \cite{cowmck77})
\begin{eqnarray}
q_{\scrm{sat}} & = & 0.2\, F_{\scrm{e}}\, kT_{\scrm{e}} \nonumber \\
 & = & 5.42 \times 10^{-3} \;
 T_7^{3/2} \; n_{0.01} \;\;  \mbox{erg cm$^{-2}$ s$^{-1}$}  
\label{eq:q_sat}
\end{eqnarray}
where $F_{\scrm{e}} = n_{\scrm{e}}
(8\,kT_{\scrm{e}}/\pi\,m_{\scrm{e}})^{1/2}$   
is the flux of hot electrons (e.g., Spitzer \cite{spit78}), $T_7$ is
the electron temperature in units of $10^7$ K, and $n_{0.01}$ is the
electron density in units of 0.01 cm$^{-3}$. 
In a gas with a cosmic Helium abundance, 2.3 particles (1.1
nuclei and 1.2 electrons) have to be evaporated (and thus heated to
temperature $T_e$) per Hydrogen atom. Thus, the evaporative flux is
\[ \frac{0.2\, F_e\, kT_e}{2.3 \times 3\,kT_e/2} = 0.058 \, F_e \;\;
\mbox{atoms cm$^{-2}$ \s.}  \]
Inserting $n_e = 0.03$ cm$^{-3}$ and $T_e = 7 \times 10^6$ K
(Trinchieri et al.\ 1997), we find that in one core crossing time 
of $3\times 10^7$ yr, all clouds with column densities $N_H \la 4
\times 10^{21}$ cm$^{-2}$ will have been evaporated. 

This limiting column density of surviving clouds is a few times smaller
than the one found by de Jong et al.\ (1990) for the
case of NGC~4696, the central cD galaxy in the Centaurus cluster which
features dusty filaments in its central regions, with a
morphology very similar to the ones in NGC~5846. As de Jong et al.\
showed, this limiting column density is still lower than that of
the ``marginally stable clouds'' of Spitzer (\cite{spit68}), which
represent clouds whose mass and central density is not quite high
enough so as to be destroyed on a free-fall timescale. 
In conclusion, there is an appreciable range in column density left for
gas cloudlets in NGC~5846 to have survived until now, but not 
gravitationally collapsed. 

\subsubsection{Energy budget for the dust lane}

In this section we investigate whether the energy lost by the hot gas
in NGC~5846 through heating dust grains and evaporating cool clouds
in the central 15$''$ can be balanced by transport of heat from the outer
regions of the hot gas. 

Heating of dust grains by hot electrons is a major heat sink of the
hot gas. Using the radial grain temperature curves in Sect.~\ref{t_d},
the {\sl IRAS\/} 60 and 100 $\mu$m flux densities of the dust features
due to heating by hot electrons alone are found to be $S_{60}$ = 3.1 mJy
and $S_{100}$ = 10.5 mJy. Using the usual formula for the total
far-infrared luminosity from Lonsdale \etal (\cite{jiswg86}), 
\[ L_{\scrm{FIR}} = 3.89 \times 10^{2}\; (2.58\, S_{60} + S_{100}) \; 
   D^2 \;  \mbox{\Lzon} \]
where $S_{60}$ and $S_{100}$ are flux densities in mJy and $D$ is the
distance in Mpc, we obtain an energy loss due to grain
heating of $7 \times 10^6$ \Lzon (cf.\ Eq.\ \ref{eq:irasflux}). 
The hot gas also looses energy by evaporating clouds, at a rate 
\[ \frac{\dot{M}}{\mu m_H} \; \times 2.3 \times \frac{3\, kT_e}{2} \]
For $0.1 < \dot{M} \;\mbox{[\Mzon\ yr$^{-1}$]}  < 0.7$, this results
in  $(2 - 13) \times 10^6$ \Lzon.  
Thus, the total energy loss of the hot gas in the central 2.3 kpc
($\cor 15''$) radius of NGC~5846 is $\sim$\,$(1 - 2) \times 10^7$ \Lzon. 

How does this compare with the energy transport into the core of
NGC~5846\,? The amount of heat transported by electron conduction is
\begin{equation}
\Gamma_{\scrm{cond}} = c \, T^{5/2} \; 4\pi R^2 \; dT/dR
\end{equation}
where $c = 6 \times 10^{-7}$ \ergcms\ K$^{-3.5}$ (e.g., McCray \& Snow
\cite{mccrsnow79}). The ROSAT data (Trinchieri \etal 
1997) show that the electron temperature drops from ($1.0 \pm 0.2$) keV
to ($0.56 \pm 0.06$) keV over the inner 2\farcm5. Hence, we find that
$\Gamma_{\scrm{cond}} = 7.5 \times 10^8$ \Lzon\ for the whole 2\farcm5
region, and $\Gamma_{\scrm{cond}} =  7.5 \times 10^6$ \Lzon\ for the
central region of 2.3 kpc radius. The uncertainty of these numbers
(due to those of the X-ray temperatures) amounts to $\sim \pm 60$\%. 

Kinetic energy of the small infalling galaxy is another ---potentially
important--- heating source for the hot gas (cf.\ Miller
\cite{mill86}; Yahil \& Ostriker \cite{yahost73}). The loss
of kinetic energy can be estimated as 
\[ \Gamma_{\scrm{kin}} = 1/2\, M_{\scrm{gal}} \,v^2 /
   t_{\scrm{infall}} \] 
Inserting $M_{\scrm{gal}} = 10^8$ \Mzon\ (a small irregular galaxy),
$v = 150$ \kms, and $t_{\scrm{infall}} = 3 \times 10^7$ yr, we find 
$\Gamma_{\scrm{kin}} = 1.2 \times 10^7$ \Lzon. 
It seems therefore that the energy losses of the hot gas are quite
adequately balanced by heat transport, especially since we have
neglected any other heating sources that may be at work, e.g., a
modest amount of star formation which may be going on. For instance,
the absence of a colour gradient in the inner regions (cf.\ Sect.\
\ref{sbprof}; Forbes et al.\ 1997) seems to indicate the presence of
an age gradient in the sense that the stars are younger in the inner
regions than they are outside, because the radial gradient of the Mg$_2$
absorption-line index does indicate a significant metallicity gradient
(Davidge \cite{davi91}). 

\subsection{Mass and excitation of nebular emission-line gas}

The main properties of the emission-line gas of NGC~5846 have been
discussed already in TdSA, so that we limit ourselves here to a
reconsideration of its properties in the light of the associated dust
absorption. 

In a situation where gas clouds are continuously evaporating, it is
quite plausible that the excitation of the ionized gas is (at least
partly) provided by collisions with hot electrons. In this respect it
is interesting to compare the heat input from electron conduction with
the observed energy emitted by the dusty filaments. 

The saturated heat flux (eq.\ [\ref{eq:q_sat}]) amounts to 0.012
\ergcms\ multiplied by the surface area of the filament
system. Assuming that the total surface area is three times the
projected surface area 
(corresponding to a 3-D morphology intermediate between a sheet and a
sphere), we obtain a total heat flow energy of $8.6 \times 10^{42}$ \ergs. 
The far-infrared luminosity of the dust lane is $8 \times 10^{40}$
\ergs\ (cf.\ Sect.\ \ref{t_d}), while the total \Ha+\NII\ luminosity
from NGC~5846 is $6.7 \times 10^{40}$ \ergs\ (TdSA). For gas temperatures
of order 10$^4$ K, the cooling curve of Raymond, Cox \& Smith
(\cite{raym+76}) indicates that the total radiated emission-line
luminosity is $\sim$\,100 times that in \Ha+\NII\ alone (cf.\ also
Ferland \& Netzer \cite{fernet83}). Thus, the
total radiated energy from the emission-line filaments is $\sim$\,$7
\times 10^{42}$ \ergs. This is strikingly similar to the heat input from
electron conduction, which makes this an energetically attractive
excitation mechanism for explaining the nebular emission in NGC~5846
(and other X-ray bright early-type galaxies, see e.g., Sparks \etal
\cite{spa+89}; Macchetto \etal \cite{duccio+96}). Detailed model
computations of the emitted emission-line intensity ratios
still need to be carried out, however. 

Another viable excitation mechanism of gas in ellipticals is
photoionization by hot post-AGB stars within the old stellar
population. Binette \etal (\cite{bine+94}) recently found that
post-AGB stars generally provide sufficient ionizing radiation to
account for the observed \Ha\ equivalent widths and luminosities in a
sample of 19 elliptical galaxies. Furthermore, more recent correlation
studies on the ISM in elliptical galaxies have shown the existence of
a strong correlation between the \Ha+\NII\ luminosity and the stellar
luminosity within the emitting region, independent of the X-ray
properties of the galaxies (Macchetto \etal \cite{duccio+96};
Goudfrooij 1997), which is also quite suggestive of a stellar origin
for the ionizing photons. Binette et al.\ (1994) also showed that the
LINER-type emission-line intensity ratios, which are typical of
nebular gas in giant ellipticals (e.g., V\'eron-Cetty \& V\'eron
\cite{vv86}; Phillips \etal \cite{phil+86}; Kim \cite{kim89};
Goudfrooij \etal \cite{paul+94b}) are well fit by their
photoionization models. 
In the present case of NGC~5846, we 
have followed the prescriptions of Binette et al.\ to predict
the \Ha\ flux from photoionization by post-AGB stars. Using model
calculations of an aging starburst, Binette et al.\ estimated the
present-day specific ionizing photon luminosity for a population of
stars typically found in elliptical galaxies to be 7.3 10$^{40}$
quanta \s\ M$_{\odot}^{-1}$. The total number of ionizing photons
($Q_{\scrm{H}}$) then follows by multiplying this value with the mass
of stars inside the \Ha-emitting region. Integrating $L_B$ within the
ellipse that encompasses the emitting region, we obtain $\log
L_{B,\,<{\scrm{em}}} = 43.88$. Assuming $M/L_B = 8$ (Binette et al.\
1994), we obtain $Q_{\scrm{H}}$ = 1.1 $10^{52}$ \s. Assuming case B
recombination  and complete re-processing of the Lyman continuum
photons, we obtain a predicted \Ha\ luminosity of 1.5 10$^{40}$
\ergs. The observed \Ha\ luminosity is only slightly higher (2.0
10$^{40}$ \ergs, cf.\ TdSA), so that post-AGB stars should also be
considered as serious candidates for providing a significant part of
the ionizing photons. However, the calculations by Binette et al.\
assumed the covering factor of the gas to be unity, which is likely to
be an overestimate since the gas distribution is much more complex
than that of the stellar component, i.e., the gas is not likely to
intercept all ionizing photons from PAGB stars within the region
outlined by the ellipse that encompasses the emitting region. This
suggests that not all of the nebular emission observed in NGC~5846 is
due to photoionization by post-AGB stars.  

The mass of emission-line gas in (elliptical) galaxies can in
principle be estimated from \Ha\ imaging data in case the
effective volume occupied by the gas is known. Since this is not well
constrained, mass estimates rely on the assumption that homogeneous
physical conditions prevail in the gas clouds. The electron density is
usually taken to be $\sim$\,10$^3$ cm$^{-3}$, appropriate for the
observed \SII\,6716/6731 doublet ratio in the nuclei of elliptical
galaxies (e.g., Phillips \etal \cite{phil+86}; Kim \cite{kim89};
Goudfrooij \etal \cite{paul+94b}). However, long-slit spectra of
{\em extended\/} gas in elliptical galaxies (Goudfrooij, in
preparation) reveal \SII\,6716/6731 doublet ratios that are in the
low-density limit (i.e., $n_e < 100$ cm$^{-3}$; cf.\ Osterbrock
\cite{oste74}), which implies that the usual mass estimates are lower
limits. In the case of X-ray bright ellipticals such as NGC~5846, one
can provide a further constraint on the mass of nebular gas by assuming 
pressure equilibrium with the hot gas, i.e.,
$n_{e,\,X} \, T_{e,\,X} = n_{e,\,{\scrm H}\alpha} \, T_{e,\,{\scrm
H}\alpha}$. Since $T_{e,\,X}/T_{e,\,{\scrm H}\alpha} \sim 10^3$, the
nebular density can be derived from the density of the hot gas. Using
the radial X-ray gas density profile of Trinchieri \etal (1997), we
obtain $n_{e,\,{\scrm H}\alpha} \sim 30$ cm$^{-3}$ within $r = 15''$,
and $n_{e,\,{\scrm H}\alpha} \sim 9$ cm$^{-3}$ at the outskirts of
the observed \Ha+\NII\ emission ($r \sim 30''$). Under the pressure
equilibrium assumption, the total mass of nebular gas is
3.0 10$^6$ \Mzon\ (a factor 54 higher than that obtained by using the
usual assumption of $n_e \sim 10^3$ cm$^{-3}$). This is similar to the
mass of neutral gas in NGC~5846 (assuming a reasonable gas-to-dust
mass ratio), which implies that the ionized fraction of the gas in
ellipticals such as NGC~5846 could well be much higher
than that usually assumed for ellipticals. If so, the amount of
ionized gas which is seen depends critically on the amount of gas
present (i.e., not merely on the number of ionizing photons available).

\subsection{Concluding remarks}
\label{concl}

The main conclusions emerging from this study are as follows:

\begin{itemize}
\item
A filamentary dust lane with a dust mass of $\sim$\,7  $10^3$ \Mzon\
has been detected in the central few kpc of NGC 5846. 
The optical extinction properties of the dust features (which are due
to the ``large'' dust grains with radius $\sim$\,0.1 $\mu$m) are
consistent with those of dust in our Galaxy. We do not expect this to
be also true for the ``small'' grains, as they are very quickly
destroyed by sputtering in the high pressure, X-ray-emitting gas pervading
NGC~5846. This can be tested by determining the extinction properties
in the UV.  The morphology of the dust features
are strikingly similar to that observed for the optical nebulosity
{\it and\/} the X-ray emission. A physical connection between the
different phases of the ISM therefore seems likely. 
In view of the combination of the filamentary morphology of the dust
lane, its close association with the ionized nebulosity, and the 
inprobability of forming dust in cooling flow condensations, we 
conclude that the dust as well as the ionized nebulosity 
most likely originate from a recent interaction with a small,
relatively gas-rich galaxy, probably of Magellanic type. 
\item
The lifetime of dust grains in the centre of NGC~5846 is 
shorter than the crossing time of a galaxy through the central 5 kpc
of NGC~5846 (where the dust is located), which indicates that the dust
must be replenished. To be consistent with the 
observed dust mass, the replenishment rate is $\sim$\,10$^{-3}$ \Mzon\
yr$^{-1}$. We argue that this replenishment can be achieved by evaporation of
cool, dense gas cloudlets that were brought in during the
interaction. The mass and density of these cloudlets is large enough to have
survived evaporation until this time, but does not have to be 
so large that they are unstable against gravitational collapse and
subsequent star formation. 
\item
The energy lost by the hot gas through heating of dust grains and
evaporation of cool gas clouds in the central few kpc of NGC~5846 is
adequately balanced by heat sources:\ 
heat transport by electron conduction into the core of the X-ray-emitting
gas and loss of kinetic energy of the infalling galaxy.
There does not seem to be a need to invoke a ``cooling flow'' 
to explain the 
X-ray observations of NGC~5846 (although we cannot firmly exclude
that some part of the radial temperature gradient in NGC~5846 is due
to cooling of hot gas). Detailed studies of the
multi-phase ISM in other X-ray bright early-type galaxies 
should enable one to assess the general
applicability of this scenario to X-ray bright ellipticals (that are
{\it not\/} at the centres of clusters). This issue will be addressed
in future papers in this series.  
\end{itemize}

\acknowledgements{We thank Christian Henkel for a useful
discussion, and the referee for a critical reading of the
manuscript. This work was initiated while PG was at the European 
Southern Observatory under Service Contract No.\
46975/DMD/96/6661/GWI, the receipt of which is hereby acknowledged.  
PG is grateful to Phil Crane for the temporary use of his office which
allowed us to conduct the data reduction for this project during
psychologically hard times. 
We have made use of the NASA/IPAC Extragalactic Database 
(NED) which is operated by the Jet Propulsion Laboratory, Caltech, 
under contract with the National Aeronautics and Space Administration.
}

\end{document}